\documentclass[preprint2]{aastex}
\shorttitle{Gyro-orbit size and the brightness temperature limit}
\shortauthors{Singal}
\begin{document}
\title{Gyro-orbit size, brightness temperature limit and implausibility of coherent 
emission by bunching in synchrotron radio sources}
\author{Ashok K. Singal}
\affil{Astronomy and Astrophysics Division, Physical Research Laboratory,\\
Navrangpura, Ahmedabad - 380 009, India}
\email{asingal@prl.res.in} 
\begin{abstract}
We show that an upper limit on the maximum brightness temperature 
for a self-absorbed incoherent synchrotron radio source is obtained 
from the size of its gyro orbits, 
which in turn must lie well within the confines of the total source extent. 
These temperature limits are obtained without recourse to inverse Compton effects or 
the condition of equipartition of energy between magnetic fields and relativistic particles.
For radio variables, the intra-day variability (IDV) implies brightness temperatures $\sim 10^{19}$~K 
in the co-moving rest frame of the source. This, if interpreted purely due to an 
incoherent synchrotron emission, would imply gyro radii $>10^{28}$~cm, the size of the universe, 
while from the causality arguments the inferred 
maximum size of the source in such a case is $\stackrel{<}{_{\sim}} 10^{15}$ cm. 
Such high brightness temperatures are sometimes modeled in the literature as  
some coherent emission process where bunches of non-thermal particles are somehow formed 
that radiate in phase. 
We show that, unlike in case of curvature radiation models proposed in pulsars, 
in the synchrotron radiation mechanism the oppositely charged particles would 
contribute together to the coherent phenomenon without the need to form 
separate bunches of the opposite charges. At the same time we show that 
bunches would disperse over dimensions larger than a wavelength in time shorter
than the gyro orbital period ($\stackrel{<}{_{\sim}}0.1$ sec). Therefore a coherent 
emission by bunches cannot be a plausible explanation of the high brightness 
temperatures inferred in extragalactic radio sources showing variability over a few hours or longer.
\end{abstract}

\maketitle

\section{Introduction}
The observed upper limit on the radio brightness temperatures of compact self-absorbed radio sources has  
till recently been thought to be an inverse Compton limit 
$T_{\rm b}\stackrel{<}{_{\sim}}T_{\rm ic} \sim10^{11.5-12} $~K (Kellermann \& Pauliny-Toth 1969). 
The argument is that at brightness temperature $T_{\rm b} \stackrel{>}{_{\sim}}
10^{12} $~K energy losses of radiating electrons due to inverse
Compton effects become so large that these result in a rapid cooling
of the system, thereby bringing the synchrotron brightness temperature
quickly below this limit. However it has been shown (Singal 1986, 2009b) 
that it is the diamagnetic effects that limit the maximum brightness 
temperatures to the somewhat lower equipartition value, 
$T_{\rm b}\stackrel{<}{_{\sim}}T_{\rm eq}\sim 10^{11-11.5} $~K, 
which also happens to be a minimum energy configuration for the system.
Due to the diamagnetic effects the energy
in the magnetic fields cannot be less than a certain fraction of that
in the relativistic particles and then an upper limit on brightness 
temperature follows naturally. Since brightness temperatures do not ever 
exceed the equipartition value $\sim 10^{11.5} $~K, inverse Compton effects do not even enter into picture.

But brightness temperatures much larger than $10^{12} $~K have been inferred 
for the variable sources, violating the incoherent brightness temperature limit. This
excess in brightness temperatures in the case of centimetre variables has been explained 
in terms of a bulk relativistic motion of the emitting component (Rees 1966,1967; Woltjer 1966), 
while in the case of metre variables this excess is explained in terms of slow 
interstellar scintillation (refraction) effects (Shapirovskaya 1982; Rickett et al. 1984). 
The relativistic Doppler factors required to explain the excess in
temperatures were initially thought to be $\sim 5 - 10$, similar to
the ones required for explaining the superluminal velocities seen in
some compact radio sources (Cohen et al. 1971, 1977; Whitney et al. 1971).  
Singal \& Gopal-krishna (1985) and Readhead (1994) pointed out that 
under the conditions of equipartition of energy between magnetic fields and 
relativistic particles in a synchrotron radio source, much higher Doppler 
factors are needed to successfully explain the variability events. 
However, the IDV implying temperatures up to $\sim 10^{19}$~K 
raises uncomfortable theoretical questions (Quirrenbach et al. 1992; Wagner \& Witzel 1995),  
as Doppler factors needed to explain observations seem too high ($>10^2$). An attempt 
has been made to explain IDV by the interstellar scintillations 
(Wambsganss et al. 1989). But close correlations between radio and optical variations 
seen in some cases rule out that all the rapid radio variations are due to interstellar 
scattering (Wagner \& Witzel 1995). One alternative could be the coherence emission models 
(see Benford \& Lesch 1998 and the references therein).
High brightness temperatures for IDV's, like 
in case of curvature radiation models proposed for pulsars, could be achieved if  
bunches of non-thermal particles are somehow formed 
that radiate in phase giving rise to coherent synchrotron emission.

In this paper we show that there is a relation between the maximum possible brightness temperature 
and the gyro radii of radiating electrons and that an upper limit on the maximum brightness 
temperature is obtained for a self-absorbed incoherent synchrotron radio source 
from a limit on the size of its gyro orbits.
Even though it may appear to give a more conservative limit on the brightness temperature
as compared to the limit of $\sim 10^{12} $~K derived for
Inverse Compton losses or due to the equipartition condition, all the same it
is a more robust value as it involves no further assumptions about the source 
than just the applicability of the standard synchrotron theory.

We also show that, unlike in case of curvature radiation models proposed in pulsars, 
in the synchrotron radiation mechanism the oppositely charged particles would 
contribute together to the coherent phenomenon without the need to form 
separate bunches of opposite charges. However we also show that 
bunches would get dispersed over dimensions larger than a wavelength in times much shorter
than the gyro orbital period and that a coherent emission by bunches cannot be a 
plausible explanation of the high brightness temperatures seen in extragalactic 
radio sources which show variability over periods of a few hours or more.

Unless otherwise specified we have used cgs system of units throughout.
\section{Gyro-orbit size and the brightness temperature limit}
A relativistic electron of rest mass $m_{0}$ and a Lorentz factor $\gamma$, gyrating in a uniform 
magnetic field $B$ with a gyro frequency $\nu_{\rm g}= \omega_g/2\pi=e\, B/(2\pi\gamma m_{0}\, c)$, 
emits most of its radiation in a frequency band
around its characteristic synchrotron frequency given by (Rybicki \& Lightman 1979),
\begin{equation}
\nu_{\rm c} = \frac {3}{2}\nu_{\rm g}\;\gamma^{3} \sin\theta = 
\frac {3}{4 \pi}\frac{e\, B_\perp}{m_{0}\, c}\;\gamma^{2}
\approx{4.2 \times 10^6}\,B_\perp \gamma^{2}.
\end{equation}
where $\theta$ is the angle between the magnetic field and the charge velocity while $B_\perp=B\sin\theta$ is the perpendicular 
field component. 
Accordingly, with the assumption that almost all the radiation from an electron is at a frequency 
$\nu \sim \nu_{\rm c}$, we can write, 
\begin{equation}
\gamma\approx \frac{1} {2 \times 10^3}\, \left({\frac {\nu}{B_\perp}}\right)^{1/2}.
\end{equation}
Gyro radius of the relativistic electron's orbit is,
\begin{equation}
\rho = \frac{c\,\sin\theta}{2\pi\nu_{\rm g}}  =\frac {3c}{4 \pi\nu} \,\gamma^{3}\;\sin^{2}\theta,
\end{equation}
where equation (1) is used to express $\nu_{\rm g}$ in terms of $\nu$.
\begin{equation}
\rho =\frac {3\lambda}{4 \pi}\,\gamma^{3}\;\sin^{2}\theta.
\end{equation}
Here $\lambda$ is the typical wavelength of radiation from an electron of Lorentz factor $\gamma$. 

An expression for the maximum brightness temperature can be found from the detailed synchrotron theory.
We assume the specific intensity $I_{\nu}$ has a usual power law in the optically thin part of the spectrum, i.e., 
$I_{\nu} \propto \nu^{-\alpha}\; {\rm or}\; \lambda^{\alpha} $, arising from a power law energy distribution of 
radiating electrons $N(E)=N_{0} E^{-s}$, with $s = 2\alpha+1$. For definiteness we shall assume $s = 3$, 
corresponding to spectral index $\alpha = 1$. For simplicity we also assume a uniform magnetic field $B$.
But none of the results derived here depend critically on these assumptions.

The maximum brightness temperature in the literature is usually considered to be at the turnover point in the 
synchrotron spectrum where the intensity peaks and that usually lies in the optically thin part of the 
spectrum (optical depth $\tau \stackrel{<}{_{\sim}}1$). But as shown by Singal (2009a,b) peak of the brightness
temperature actually lies deep within the optically thick region ($\tau \sim 3$). 

The specific intensity in a synchrotron self-absorbed source is given by,
\begin{eqnarray}
I_{\nu} = c_{14}^{-1}(s)\, \nu^{5/2} \:(B_\perp) ^{-1/2}\,[1-\exp(-\tau)], 
\end{eqnarray} 
where $c_{14}(s)$ values are tabulated in Pacholczyk (1970). 

Using Rayleigh-Jeans law for the specific intensity,
\begin{eqnarray}
I_{\nu} = \frac{2\:k\: T\, \nu^2}{c^2}= 3 \times 10^{-37}\: T\, \nu^{2},
\end{eqnarray} 
we can determine the maximum brightness temperature (for $s = 3$) from Singal (2009a,b) as
\begin{equation}
T_{\rm m} \approx 0.5 \times  10^6 \left({\frac {\nu_{\rm m}}{B_\perp}}\right)^{1/2}.
\end{equation}

Using equation 2 we can write $T_{\rm m}$ in terms of the typical Lorentz factor $\gamma_{\rm m}$ of the electrons 
radiating at the maximum brightness temperature as
\begin{equation}
T_{\rm m} \approx 10^{9}\,\gamma_{\rm m} ,
\end{equation}
which is in conformity with the fact that 
in an incoherent emission, where relativistic particles radiate independently 
of each other, thermodynamics constrains the brightness temperature $T_{\rm b}$ to be
$kT_{\rm b}\stackrel{<}{_{\sim}}k T_{\rm e}/3$ where $k=1.38 \times 10^{-16}$ erg/deg is Boltzmann constant 
and $T_{\rm e}=\gamma\,m_{0}\: c^{2}/k = 5.93 \times 10^{9}\,\gamma$ is kinetic temperature of the 
radiating particles (Scheuer and Williams 1968; Altschuler 1989), implying thereby, 
$T_{\rm b}\stackrel{<}{_{\sim}} 2 \times 10^{9}\gamma$.

Now  from equations 4 and 8 we get the gyro-orbit size as,
\begin{equation}
\rho_{\rm m} \sim \frac {3\lambda_{\rm m}}{4 \pi}   \left(\frac{T_{\rm m}}{10^{9}}\right)^{3}\sin^{2}\theta.
\end{equation}
Using average value of $\sin^{2}\theta=2/3$, we can write,
\begin{equation}
{2 \pi}\rho_{\rm m} \sim  {\lambda_{\rm m}}   \left(\frac{T_{\rm m}}{10^{9}}\right)^{3}.
\end{equation}

Thus an upper limit to the brightness temperature at an observing wavelength $\lambda$ can be found if the gyro-orbit 
size of the corresponding radiating electrons is constrained, e.g., from an upper limit on the inferred size of 
the radiating source component. 

Now if try to explain the observed radio variability data of AGNs in terms of purely an 
incoherent synchrotron radiation, we get some very unexpected results. 
For the IDV at GHz frequencies, the inferred brightness temperatures 
turn out to be $T_{\rm b}\sim 10^{19}$~K (Quirrenbach et al. 1992), implying 
$\rho> 10^{28}$~cm, the size of the known universe. 
We can understand it also like this. A  brightness temperature
$T_{\rm b}\sim 10^{19}$~K could be achieved only if the kinetic temperature of the 
radiating particles exceeds this value, implying $\gamma >10^{9}$. Also the relation 
between frequency and gyro-frequency $\nu \sim \nu_{\rm g}\;\gamma^{3}$ can be translated 
into that between wavelength and gyro-radius as $\lambda \sim \rho\;\gamma^{-3}$ 
or $\rho \sim \lambda\;\gamma^{3}$. Therefore 
for GHz frequencies ($\lambda \sim 10$ cm) we get $\rho> 10^{28}$~cm.
The source size on the other hand cannot be much larger than $\sim 10^{14.5 - 15}$~cm 
from the causality arguments for these variability events, 
and thus their gyro-orbit sizes also cannot exceed  this value (a part cannot be larger than the 
whole!). This fact alone makes it imperative that some extraneous factor like relativistic beaming 
or interstellar refraction or some coherent phenomena are invoked to explain these high brightness temperatures.

From the size limits of $\sim 10^{14.5-15}$~cm on the diameter of the gyro orbit, we immediately 
get a gyro limit on the brightness temperature $T_{\rm g}\sim 10^{13.5}$~K (equation 10).
This gyro temperature limit is arrived at using just the applicability of the standard
synchrotron theory, without any a-priori assumption about the equipartition 
conditions in the source or even the inverse Compton effects. 
Even in the static compact synchrotron sources, where no variability is observed, 
the gyro-orbit diameter has to be still smaller than the
observed source size. Actually  the gyro-orbit diameter 
will have to be very much smaller than the observed source size. It is not just a question of 
``a part being not larger than the 
whole'', in order to apply the self-absorbed synchrotron theory, one has to assume a large 
number of synchrotron orbits along the line of sight within the total source extent, for the
optical depth to exceed unity. Therefore the typical gyro orbit has to be perhaps many orders 
of magnitude smaller than the total source extent, and the gyro temperature limit $T_{\rm g}$ will then 
move close to $\sim 10^{12}$~K.

The arguments usually offered in the literature against very high $T_{\rm b}$ values are based on the inverse Compton 
catastrophe where an incoherent synchrotron source cannot sustain these for long as the extremely large inverse 
Compton losses would very rapidly bring them to the incoherent brightness temperature limit. But from the gyro-size 
arguments we can say that an incoherent synchrotron source in the first place could not have ever achieved such 
extremely high brightness temperatures. Even though the derived temperatures 
limits may appear somewhat conservative than those derived from Inverse Compton effects or
the equipartition conditions, but the values derived here much more robust as these involve
no further assumptions about the source than just the applicability of the standard
synchrotron theory.
\section{Estimates of gyro-orbit sizes, number densities and other parameters in compact synchrotron sources} 
We can reverse the roles and employ the observed maximum $T_{\rm b}$ values along with 
equations 7, 8 and 10 to estimate gyro-orbit sizes and some typical numbers 
inside compact synchrotron sources. 
For the brightness temperature seen in VLBI observations, 
$T_{\rm b}\stackrel{<}{_{\sim}} 10^{11.5}$~K (Kellermann \& Pauliny-Toth 1969). Then we get a gyro radius 
$\rho \sim 10^{8.5}$ cm, implying a gyro orbital period ${\cal T}\stackrel{<}{_{\sim}}0.1$ sec, 
for the electrons radiating near the turnover frequency $\nu \sim 10^{9}$~Hz. This gives 
$\gamma \sim 10^{2.5}$ and $B\sim 10^{-2.5}$ Gauss. 

We can make an order of magnitude estimate of the number density of particles in the source, by noting that 
near the spectral turnover, where $T_{\rm b} \sim 10^{11.5} $~K, there is an equipartition of energy between 
particles and magnetic fields (Singal 1986, 2009b). Also this is the region where the energy density of photons, 
$W_{\rm p}$, approaches that of magnetic fields (Kellermann \& Pauliny-Toth 1969). We can estimate $W_{\rm p}$ as, 
\begin{equation}
W_{\rm p}=\frac{4 \pi}{c} \int I_{\nu}\: {\rm d}\nu 
\approx \frac{4 \pi}{c} I_{\rm m}\: \nu_{\rm m}= \frac{8 \pi}{c^3}\:k T_{\rm m}\:\nu_{\rm m}^{3}.
\end{equation} 
Thus one can get the number density of radiating charges 
$N m c^2 \gamma \approx W_{\rm p} \approx 8 \pi k T_{\rm m}/ \lambda_{\rm m}^{3}$, and with 
$3kT_{\rm m}\sim kT_{\rm e}= m c^2 \gamma$, we get $N \sim 8 \pi/3\lambda_{\rm m}^3$. 
For $\nu_{\rm m}\sim 1$ GHz, we get a typical number density $\sim 10^{-4}$ to $10^{-3}$ cm$^{-3}$.

\section{Implausibility of coherent emission by bunches in synchrotron radio sources} 
It has been shown that synchrotron radiation process 
does not allow MASER type coherent emission (Pacholczyk 1970; Rybicki \& Lightman 1979).
Alternate coherence emission models have been proposed in the literature 
(Cocke \& Pacholczyk 1975; Cocke et al. 1978; Colgate \& Petschek 1978; 
Benford \& Lesch 1998 and the references therein). 
Coherent emission could be achieved if bunches of non-thermal particles are somehow 
formed, which radiate in phase through antenna mechanism. As an example, 
in the case of pulsar magnetosphere there are models of curvature radiation where
coherence emission by bunches has been proposed to explain extremely high brightness temperatures 
observed in pulsars. Could similar models succeed in the case of synchrotron radiation in IDV's?   

Now in an overall neutral plasma there may be 
oppositely charged particles, e.g., a pair plasma comprising electrons and positrons. 
The contributions of electrons and positrons are indistinguishable in incoherent synchrotron cases. 
However, it is important to know if opposite charges occupying the same phase space 
cancel each other's contribution to the synchrotron radiation?
In the pulsar models, electrons and positrons can cancel each others radiation 
fields. The biggest hurdle in those models is to form stable bunches of sizes less than a wavelength 
for each type of charges, with bunches of opposite charges separated from each other by more than a 
wavelength (Melrose 1992). Not only such bunches of the like-charges may be unstable but strong 
electrostatic attraction between oppositely charged bunches tends to hinder the formation of such bunches. 
But it turns out that in the synchrotron case one need not separate the positive and negative
charges in an overall neutral plasma. The radiation fields of electrons and positrons lying in the same 
phase space do not cancel each other.

The radiation field for a moving charge, derived from the Li\'{e}nard-Wiechert potentials, is given by (Jackson 1975), 
\begin{equation}
{\bf E}=\frac {e}{c} \frac{{\bf n}\times\{({\bf n}-\mbox{\boldmath $\beta$})\times
\dot{\mbox{\boldmath $\beta$}}\}}{R(1-\mbox{\boldmath $\beta$}.{\bf n})^{3}} \;.
\end{equation}
Equation 12 can be derived also from first principles using Coulomb's law and the 
Lorentz transformations including relativistic Doppler factors (Singal 2011), 
without going through Li\'{e}nard-Wiechert potentials.
In the case of gyration in a magnetic field {\bf B}, the acceleration is given by 
$m c  \dot{\mbox{\boldmath $\beta$}}\gamma = e {\mbox{\boldmath $\beta$}} \times {\bf B}$,
from which we get,
\begin{displaymath}
{\bf E}=\frac {e^2}{m c^2 \gamma} 
\frac{{\bf n}\times\{({\bf n}-\mbox{\boldmath $\beta$})\times
({\mbox{\boldmath $\beta$}} \times {\bf B})\}}{R(1-\mbox{\boldmath $\beta$}.
{\bf n})^{3}} \;,
\end{displaymath}
which is independent of the sign of the charge.

Thus charges of opposite sign lying in the same phase space, i.e. moving together with the same 
velocity \mbox{\boldmath $\beta$}, will have opposite 
acceleration vectors in the magnetic field $\bf B$ and consequently give rise to similar electric fields 
at the observer towards direction {\bf n}. The radiation fields from electrons and positrons within a bunch  
will augment each other, and bunches of oppositely charged particles separated from each other by more than 
a wavelength are not necessary for coherence purpose. 
In the case of curvature radiation models, both type of charges undergo similar acceleration as they follow the 
curvature of the magnetic field lines and their radiation fields, being in opposite directions, get cancelled. 
But in the synchrotron case the accelerations are of opposite signs, resulting 
in electric fields being in the same direction even from opposite charges. Thus a very big hurdle, viz. forming and 
maintaining separate bunches of oppositely charged particles, faced in the curvature radiation models, can be  
avoided in the coherence emission models in the synchrotron case. 

In a synchrotron source, all electrons gyrate in a clock-wise fashion (looking along the magnetic 
field vector). Now radiation at an observational frequency $\nu$ arises mostly from electrons 
with $\gamma$ given by equation 2 and the radiated power lies in a cone of opening angle 
$\sim 1/\gamma$. Thus only charges contributing effectively to the radiation at any instant are 
the ones that are moving within a narrow angle 
$\psi \sim 1/\gamma$ with respect to the line of sight to the observer. 

During every gyration cycle an electron radiates towards the observer for a time interval 
$\sim 2/(\omega_g \gamma\sin\theta)$, but due to Doppler effect, the radiation to the observer 
appears as a pulse of duration $ \Delta t \sim 1/(\omega_g \gamma^3 \sin\theta) 
\sim 1/\omega_c$ (Rybicki \& Lightman 1979). Then the pulse width in space will be 
of length $\sim \lambda$ and the average electric field stronger by a factor $\sim \gamma^3$
within the pulse window than without. Thus two electrons moving towards the observer with a projected distance 
$d$ apart will have their pulses arriving at the observer separated by a time 
interval $\sim d/c$. Thus the pulse windows will overlap, with the Fourier components adding in phase,
only if the electrons are less than a projected distance $\lambda$ of each other. 
However if the two charges are separated by a distance more than $\lambda$ 
then there will not be any overlap in their main pulse windows, and their fields will not add in phase. 
Thus for a coherence to occur, the charges have to form a bunch lying within a region of length $\sim \lambda$ 
along the line of sight and thereby radiating in phase towards the observer. 
The lateral width $W$ of the coherence volume for charges radiating in a cone of opening angle 
$\psi \sim 1/\gamma$, can be calculated from the condition $W \psi \sim \lambda$ or 
$W \sim \gamma \lambda$, which gives a lateral cross-section $\sim \pi \gamma^2 \lambda^2$. 
The coherence volume will be in the shape of a chapati \footnote{A thin flat circular unleavened Indian bread}, 
with a thickness $\lambda$ and a lateral cross-section $\pi \gamma^2 \lambda^2$. 
This gives us a total coherence volume $V_c \sim \pi \gamma^2 \lambda^3$. 
A somewhat more rigorous approach yields $V_c \sim \gamma^2 \lambda^3/\pi$ (Melrose 1992). 
Thus for coherence emission to take place there has to be a bunch of charged particles in the coherence volume,
and for every gyro period ${\cal T}$ this bunch will radiate towards the observer coherently for a time $\sim {\cal T}/\gamma^3$.

It should be noted that an electron and positron pair emitting coherently is not 
co-located with concentric gyro orbits, since while spiraling in opposite sense, whenever the pair crosses each other,    
their velocity vectors will be pointing in opposite directions and the two will never be in the same phase space. 
For coherence emission, both type of charges must be at the some location, having similar velocity vectors, and thus 
radiating in the same direction (Like two circles touching each other externally with a common tangent). 
Now pairs of such bunches of oppositely charged particles, 
will each radiate towards the observer for a short duration $\sim {\cal T}/ \gamma^3)$, 
and after spiraling in opposite directions will again come together after the gyro period ${\cal T}$,  
radiating in the same direction and thus becoming part of the same bunch radiating 
coherently as far as the observer is concerned. 

A bunch of charges may radiate coherently only as long as the bunch stays together within dimensions less than a 
wavelength. In that case the observed coherent emission could attain brightness temperatures much above 
the theoretical limit for the incoherent synchrotron emission. The coherent emission will last as long 
as the bunch survives. But as we discuss below there are serious difficulties with the stability 
of the bunches. The bunches are too short-lived and that their coherent emission at most may last over time 
intervals only of the order of a gyro-orbit period ${\cal T}$ ($\stackrel{<}{_{\sim}}0.1$ sec, Section 3). 
On the other hand the variable extragalactic sources 
show IDV periods of a few hours. Any coherence phenomenon  
{\em to explain these variabilities}, must also last over similar time scales. 
However, as we argue below, a bunch, where a number of charges moving together occupy a region smaller 
than a wavelength, cannot be a stable configuration over such time scales in a synchrotron case. 

A tiny velocity dispersion, implying small spread in Lorentz factor $\gamma$, would make the bunch 
disperse in a time interval much smaller than the gyro-orbit period. The gyro frequency 
$\nu_{\rm g}= e\, B/(2\pi\gamma m_{0}\, c)$ is inversely proportional to $\gamma$. 
Then two charges with a difference $\Delta\gamma$ in Lorentz factors will differ in their gyro period ${\cal T}$ by, 
\begin{equation}
\frac{\Delta {\cal T}}{{\cal T}}=-\frac{\Delta\nu_{\rm g}}{\nu_{\rm g}}=\frac{\Delta\gamma}{\gamma}.
\end{equation}
Thus two charges, starting from the same location, will be about a distance $\Delta x=c \Delta {\cal T}$ apart after one gyro orbit.
This will result in charges in a bunch getting spread over regions $\Delta x=c {\cal T} \Delta\gamma/\gamma$ in time ${\cal T}$. 
As $c {\cal T}=2\pi\rho \approx\lambda \gamma^3$ (equation~3, with average value of $\sin^{2}\theta=2/3$), it means $\Delta x \approx \lambda\gamma^2 \Delta\gamma$.
It can be written in terms of the velocity dispersion $\Delta \beta$, as for relativistic particles ($\beta \sim 1$), 
$\Delta \gamma =\gamma^3 \Delta \beta $. Thus due to dispersion $\Delta \beta$ in velocity of the charges within the bunch, 
their  gyro-orbit sizes also have a dispersion $\Delta x \approx \lambda  \gamma^5 \Delta \beta$. 
As for the relativistic particles emitting synchrotron emission, 
$\gamma \gg1$, we see that just within a gyro orbital period a bunch will get spread over regions 
larger than $\lambda$, even for $\Delta \beta$ as low as $10^{-12}$, i.e., for velocity dispersion of the order 
only a fraction of a mm/sec.

Normally a bunch with a velocity dispersion $\Delta \beta$ would get spread over a region $c t\Delta \beta $ 
in a time period $t$ and for the coherence to be effective, particles in a bunch should remain confined to within 
a wavelength $\lambda$ which would  require that that $c t\Delta \beta\stackrel{<}{_{\sim}} \lambda$. 
Now there will be $t/{\cal T}$ gyro orbits in time $t$, so from equation~13 a bunch will get spread over a region 
$c t \Delta\gamma/\gamma = c t\Delta \beta \gamma^2$ in time $t$. 
Thus due to the difference in gyro-orbit periods the bunch would get spread over a region larger by a factor $\gamma^2$ 
than due to just dispersion $\Delta \beta$ in velocity. Or in other words a bunch will get spread over the region 
$c t\Delta \beta $ in a much shorter time  $t/\gamma^2$. 
Effectively any such bunch, even if somehow formed, will get dispersed over regions larger than 
a wavelength in a gyro orbital period ${\cal T}\stackrel{<}{_{\sim}}0.1$ sec, 
thus altogether destroying or at least largely 
diminishing the coherence capability of the bunch within fraction of a second. Thus even if coherent  
emission due to bunches does give rise to high brightness temperatures, the bunches will get dispersed in a very short time 
($\stackrel{<}{_{\sim}}0.1$ sec) and the emission will drop on a similar time scale, not what is actually observed.
In other words even if theoretical models based on coherent synchrotron emission by bunches could achieve extremely high 
brightness temperatures seen in IDV's, this high brightness will be short-lived by many orders 
of magnitude than what actual observations show and it looks quite implausible that coherent emission by bunches 
could consistently account for the observed variability time scales of a few hours or more.

Since the flux-density variability could be a substantial fraction of the total 
flux-density of the source, it is essential that there will have to be 
large number of bunches spread over the source extent. 
Above discussion is applicable to each and every bunch and thus to 
the sum total of their emission in the whole source. Of course different 
bunches will be randomly located with respect to each other and thus  
would not be in phase with each other, so coherence between various 
bunches is not expected in any case. 
Could it be that there were formation of new bunches in succession in response to, 
say, some local instability, and which could then account for longer time scales of the 
observed variability? Actually each of these bunches will last for a short time interval and their statistical 
addition will still have the primary time scale of the individual bunches. Consider for example a variability 
event where the flux-density reaches some peak value before declining on the time scale of hours. 
All the bunches that may be contributing to the peak flux-density would get dispersed very fast and the flux-density 
too would then decline equally fast. Only in a very contrived situation will the individual bunches follow each other 
so systematically that the resultant variability curve will have smooth shape lasting for many order of magnitude 
larger than that of the individual bunch time scale. In general the variability curves will show the sharp time 
scales of bunches. Of course all electrons in certain region might not be contributing to the bunch, but then the coherence 
emission will be proportional to the square of the number of bunched electrons only and for which the survival of 
the bunches is necessary for a lasting coherent emission. One could envisage a combination of coherence and relativistic 
beaming, but if the dominant emission is due to coherence then all the difficulties discussed above would remain  
still valid.   

\section{Conclusions}
We have shown that there is a relation between maximum possible brightness temperature 
and the gyro radii of radiating electrons 
and that an upper limit on the maximum brightness temperature is obtained 
for a self-absorbed incoherent synchrotron radio source from the size of its gyro orbits.
Even though it may appear to give a more conservative limit on the brightness temperature
as compared to the limit of $\sim 10^{12} $~K derived for
Inverse Compton losses or due to the equipartition condition, all the same it
is a more robust value as it involves no further assumptions about the source 
than just the applicability of the standard synchrotron theory.
We have also shown that, unlike in case of curvature radiation models proposed in pulsars, 
in the synchrotron radiation mechanism the oppositely charged particles lying in the same 
phase space would contribute together to the coherent phenomenon without the need to form 
separate bunches for particles with opposite charges. However we have also shown that 
bunches are short-lived and get dispersed over dimensions larger than a wavelength in times shorter
than the gyro orbital period. Thus even if theoretically a coherent emission by bunches could attain 
high brightness temperatures seen in extragalactic radio sources, still it cannot be a 
plausible explanation of the observed variability which lasts over periods of a few hours or more.

\end{document}